\documentclass[twocolumn,showpacs,prl]{revtex4}
\usepackage{amsmath}
\usepackage{graphicx}
\usepackage{dcolumn}
\usepackage{bm}

\begin{document}

\title{Suppression of superconductivity in granular metals}
\author{I.~S.~Beloborodov}
\affiliation{Materials Science Division, Argonne National
Laboratory, Argonne, Illinois 60439}
\author{A.~V.~Lopatin}
\affiliation{Materials Science Division, Argonne National
Laboratory, Argonne, Illinois 60439}
\author{V.~M.~Vinokur}
\affiliation{Materials Science Division, Argonne National Laboratory, Argonne, Illinois
60439}
\date{\today}

\begin{abstract}
We investigate the suppression of the superconducting transition
temperature due to Coulomb repulsion in granular metallic systems
at large tunneling conductance between the grains, $g_{T}\gg 1$.
We find the correction to the superconducting transition
temperature for 3$D$ granular samples and films. We demonstrate
that depending on the parameters of superconducting grains, the
corresponding granular samples can be divided into two groups: (i)
the granular samples that belong to the first group  may have only
insulating or superconducting states at zero temperature depending
on the bare intergranular tunneling conductance $g_T,$  while (ii)
the granular samples that  belong to the second group in addition
have an intermediate metallic phase where superconductivity is
suppressed while the effects of the Coulomb blockade are not yet
strong.
\end{abstract}

\maketitle

A great deal of current experimental and theoretical research in
mesoscopic physics focuses on properties of {\it inhomogeneous}
granular superconductors~\cite{Valles,experiment,Jaeger,Simon}.
The interest is motivated not only by their unusual properties
such as negative magnetoresistance~\cite{Beloborodov99}, but, even
to a higher extent, by the fact that they represent an exemplary
disordered electronic system with experimentally accessible
tunable parameters. A specific feature of granular metallic
materials is the important role of the Coulomb interaction that
strongly affects their behavior. The interplay between the Coulomb
interaction effects and disorder was shown to suppress an electron
transport in granular metals~\cite{Efetov,Efetov02,Lopatin03}; the
fundamental question that remains open is how the Coulomb
repulsion affects the superconducting properties of granular
metals, and in particular, the superconducting transition
temperature.

Coulomb interactions decrease the critical temperature, this
effect is amplified significantly in the presence of disorder and
becomes especially pronounced in lower dimensions. For example in
$2D$ (disordered) superconducting films the effect of Coulomb
repulsions is known to noticeably decrease the critical
temperature
~\cite{Ovchinnikov73,Fukuyama81,Finkelstein87,Larkin99}. The
mechanism of suppression of the transition temperature can be
understood as a result of the renormalization of the electron
interaction in the Cooper channel by the Coulomb repulsion in the
presence of scattering by impurities~\cite{Altshuler}. Since in
granular samples both disorder and the strong Coulomb repulsion
effects present, one expects that similar mechanism governs the
reduction of the superconducting critical temperate and becomes
even more pronounced for granular superconductors.

In this Letter we investigate the influence of Coulomb repulsion
on the superconducting transition temperature of granular metals
and demonstrate that depending on the parameters of the grains,
the granular materials can be divided into two groups: (i)
materials that may be found in either insulating or
superconducting state at zero temperature and (ii) materials that
can, in addition, exhibit an intermediate metallic state. The
parameter that governs the corresponding behavior is the tunneling
conductance $g_T$. The transport properties of granular metals
were recently extensively studied in Ref.
~\cite{Efetov,Efetov02,Lopatin03}. It was shown that for 3D
samples there is a critical value of the bare tunneling
conductance~\cite{Lopatin03}
\begin{equation}
\label{gC}
 g_T^C=(1/6\pi)\,\ln(E_C/\delta),
\end{equation}
where $E_C$ is the Coulomb energy and $\delta$ is the mean level
spacing in a single grain, that allows to classify granular
samples by their transport properties: samples with $g_T<g_T^C$
are insulators at zero temperature while samples with $g_T>g_T^C$
are metals. We show that in order to classify 3D superconducting
granular samples with respect to their possible ground state at
zero temperature, an additional characteristic conductance $g_T^*$
should be introduced:
\begin{equation}
 g_T^* \approx {A \over \pi} \ln^2 (E_C/T_c^0)  \label{g_star}
\end{equation}
where $T_c^0$ is the superconducting critical temperature of a
single grain and $A$ is the numerical coefficient defined below.
We demonstrate that granular superconductors can be conveniently
classified depending on the relation between $g_T^C$ and $g_T^*$ into two
groups: \\
(i) Group A: Granular samples with $g_T^*<g_T^C$  may have only two phases
at zero  temperature depending on the tunneling conductance $g_T.$
They are either superconductors if $g_T>g_T^*,$ or insulators if $g_T<g_T^*$.
\\
(ii) Group B: Samples with  $g_T^*> g_T^C$ can be found in three
phases. They are superconductors if $g_T>g_T^*$, insulators if
$g_T<g_T^C$, or metals  in the intermediate region
$g_T^*>g_T>g_T^C$.

Thus, in the latter case there exists an intermediate metallic
region $g_T^*>g_T>g_T^C$ where the superconductivity is suppressed
by the Coulomb interaction while effects of Coulomb blockade are
not yet strong.

Presenting our final result for the suppression of the
superconducting transition temperature we distinguish the high
temperature, $T_c^0>g_T\delta$, and low temperature,
 $T_c^0<g_T\delta$, regimes:   For 3D systems the
 suppression of the transition temperature is given by
\begin{subequations}
\label{main}
\begin{equation}
\frac{\Delta T_c}{T^0_c} = - \left\{
\begin{array}{lr}
\frac{A}{\pi g_T}\ln^2 \frac{g_TE_C}{T^0_c},
\hspace{2.5cm} T_c^0 > g_T\delta &  \\
\frac{A}{\pi g_T}\left[ \ln^2 \frac{g_T E_C }{T^0_c} -
\frac{1}{2}\ln^2 \frac{g_T\delta}{T_c^0}\right], \hspace{0.3cm}
T_c^0 < g_T\delta &
\end{array}
\right. , \label{mainresult1}
\end{equation}
while for granular films we obtain
\begin{equation}
\frac{\Delta T_c}{T^0_c} = - \left\{
\begin{array}{lr}
\frac{1}{6\pi^2 g_T}\ln^3 \frac{g_TE_C}{T^0_c},
\hspace{2.3cm} T_c^0 > g_T\delta &  \\
\frac{1}{6\pi^2 g_{T}}\left[\ln^3 \frac{g_T E_C }{T^0_c} -
\frac{1}{4}\ln^3 \frac{g_T \delta }{T^0_c} \right], \hspace{0.2cm}
T_c^0 < g_T\delta  &
\end{array}
\right. . \label{mainresult2}
\end{equation}
\end{subequations}
Here $A = g_T a^3\int d^3 q/(2\pi)^3\varepsilon_{\bf q}^{-1}$ is
the dimensionless constant where $\varepsilon_{\bf q} =
2g_T\sum_{\bf a}(1-\cos{\bf qa})$ with  $\{{\bf a}\}$ being the
lattice vectors. Equations~(\ref{main}) hold at temperatures
$T^0_c - T_c \ll T^0_c$.

Turning to the discussion of the results (\ref{main}), we note
that the existence of two qualitatively different  temperature
regimes ($ T>g_T\delta$ and $T<g_T\delta$ ) is not surprising in
view of the results for corrections to conductivity obtained in
Ref.~\cite{Lopatin03}, where it was shown that the temperature
dependence of the conductivity at $T<g_T\delta$ is dominated by
the contribution from coherent electron motion at large distances,
while at $T>g_T\delta$ the conductivity behavior is controlled by
the scales of the order of the grain size. From
Eq.~(\ref{mainresult1}) one can see that the logarithmic
corrections appear even in $3D$ case; this property can serve as a
specific characteristic of granular metals. In the high
temperature regime $T>g_T\delta$ similar logarithmic corrections
(with the same argument but different powers) were found in the
corrections to the conductivity and density of states
~\cite{Efetov02,Lopatin03}. Yet, the result for the critical
temperature suppression  in the high temperature regime depends on
the dimensionality and is not universal contrary to the
logarithmic corrections to conductivity
 found in the same regime in Ref.~\cite{Efetov02,Lopatin03}.
We see that in granular superconducting films the suppression of
the transition temperature is much stronger than that in $3D$
granular samples. The extra logarithm power in two dimensions is
due to the contribution of low momenta $q\ll a^{-1}.$

Our classification for 3$D$ granular superconductors can be
derived from the following arguments: The critical conductance
$g_T^*$ is defined by the condition that the correction
(\ref{mainresult1}) becomes of the order of unity at $g_T\approx
g_T^*.$ Considering first the case where $g_T^*<g_T^C,$ we see
that the system should be a superconductor at $g_T>g_T^*$, since
the conductance renormalization due to the effect of Coulomb
interaction  $\Delta g_T$ taken at the temperature $T \sim T_c^0$
is smaller than $g_T$ for all $g_T$ larger than  $g_T^*$. Indeed
 $\Delta g_T  = (1/6\pi ) \ln(g_T E_C/T_c^0) < g_T^*.$ On the other
 hand, if $g_T<g_T^*$, then the interaction in the Cooper channel is
 completely suppressed and the system becomes an insulator at zero
 temperature because $g_T < g_T^C$. Thus in the case  $g_T^*<g_T^C$
 depending on the value of conductance $g_T$
 one observes either insulating, at $g_T<g_T^*$, or superconducting,
 at $g_T>g_T^*$, phases at zero temperature.
 The second case, where $g_T^*>g_T^C$, is qualitatively different
 due to appearance of the intermediate metallic phase:
In the region $g_T>g_T^*$ the system is a superconductor, in the
region $g_T^*>g_T>g_T^C$ the interaction in the Cooper channel is
completely suppressed, and the system is metallic since
$g_T>g_T^C$, and finally in the region $g_T<g_T^C$ the system
becomes an insulator.

Now we turn to the quantitative description of our model and
derivation of Eq.~(\ref{mainresult1}):  We consider a
$d-$dimensional array of superconducting grains in the metallic
state. The motion of electrons inside the grains is diffusive and
they can tunnel between grains. We assume that in the absence of
the Coulomb interaction, the sample  would have been a good metal
at $T > T_c$.

\begin{figure}[tbp]
\includegraphics[width=.95\linewidth]{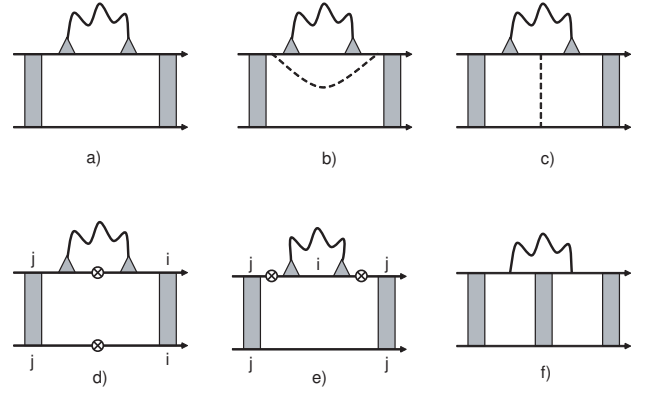}
\caption{Diagrams describing the first order self-energy
corrections, $\delta\lambda_1$ to the  Cooperon channel coupling
constant $\lambda$ of granular metals due to the electron-electron
repulsion. The solid lines denote the propagator of electrons, the
wave lines describe screened Coulomb interaction and the dashed
lines describe the elastic interaction of electrons with
impurities. The black rectangle and triangle denote the
renormalized Cooperon, see Eq.~(\ref{cooperon}), and impurity
vertex of granular metals respectively. The indices $i$ and $j$
stands for the number of the grains. The tunneling vertices are
described by the circles. All diagrams should be taken with a
factor of 2. The final result for the sum of the diagrams (a) -
(f) is given by Eq.~(\ref{af}).} \label{fig:1}
\end{figure}

The system of weakly coupled superconducting grains is described
by the Hamiltonian
\begin{subequations}
\label{hamiltonian}
\begin{equation}
\hat{H} = \hat{H}_{0} + \hat{H}_{c} + \hat{H}_{t}.
\label{hamiltonian1}
\end{equation}
The term $ \hat{H}_{0}$ in Eq.~(\ref{hamiltonian1}) describes
isolated disordered grains with an electron-phonon interaction
\begin{equation}
\hat{H}_{0} = \sum\limits_{i,k}
\epsilon_{i,k}a^{\dagger}_{i,k}a_{i,k} - \lambda
\sum\limits_{i,k,k'}
a^{\dagger}_{i,k}a^{\dagger}_{i,-k}a_{i,-k'}a_{i,k'} +
\hat{H}_{imp},
\end{equation}
where $i$ stands for the number of the grains, $k\equiv ({\bf
k},\uparrow )$, $-k \equiv (-{\bf k}, \downarrow)$; $\lambda>0$ is
the interaction constant; $a^{\dagger}_{i,k} (a_{i,k})$ are the
creation (annihilation) operators for an electron in the state $k$
of the $i$-th grain and $\hat{H}_{imp}$ describes elastic
interaction of electrons with impurities. The term $\hat{H}_{c}$
in Eq.~(\ref{hamiltonian1}) describes the Coulomb repulsion inside
and between the grains and is given by
\begin{equation}
\label{Hc}
\hat{H}_{c}={\frac{{\ e^{2}}}{{\ 2}}}\,\sum_{ij}\,\hat{n}_{i}\,C_{ij}^{-1}\,
\hat{n}_{j},
\end{equation}
where $C_{ij}$ is the capacitance matrix and $\hat{n}_{i}$ is the
operator of electrons number in the $i$-th grain. The last term in
the right hand side of Eq.~(\ref{hamiltonian1}) is the tunneling
Hamiltonian
\begin{equation}
\label{Ht} \hat{H}_{t} = \sum_{ij,p,q} t_{ij} a^{\dagger}_{i,p}
a_{j,q},
\end{equation}
\end{subequations}
where $t_{ij}$ is the tunneling matrix element corresponding to
the points of contact of $i$-th and $j-$th grains.

The transition temperature, $T_c$ of the granular metals is determined from
the pole of the superconducting propagator~\cite{AGD}. As in
the case with homogeneous superconducting metals in the absence of
the electron-electron repulsion the transition temperature of
granular metals is independent of disorder.

To study the suppression of the transition temperature in granular
metals we take into account the electron-electron repulsion in the
Cooper channel. The first order correction $\delta \lambda$ to the
Cooperon channel coupling constant $\lambda$ due to
electron-electron repulsion is given by the diagrams in Figs.~1
and 2. These diagrams contain essentially the averaged
one-particle Green functions (solid lines), effective screened
Coulomb propagators (wave lines) and dashed lines that describe
the elastic interaction of electrons with impurities. In the
regime under consideration all characteristic energies are less
than Thouless energy $E_T=D/a^2,$ where $D$ is the diffusion
coefficient of a single grain. This allows us to use the zero
dimensional approximation for a single grain diffusion and
Cooperon propagators. The electron hopping between the grains can
be included using the diagrammatic technique developed in
Refs.~\onlinecite{Beloborodov99,Efetov}. The complete expression
for the renormalized Cooperon of granular metals (black rectangle
in Fig.~1 and 2) has the following form
\begin{equation}
\label{cooperon} C(\omega_n,{\bf q}) = (2\pi \nu
\tau^2)^{-1}(|\omega_n| + \varepsilon_{\bf q}\delta)^{-1},
\end{equation}
where ${\bf q}$ is the quasi-momentum, $\omega_n = 2\pi T n$ is
the bosonic Matsubara frequency and $\nu$ is the density of states
on the Fermi surface. The parameter $\varepsilon _{\bf q}$ in the
right hand side of Eq.~(\ref{cooperon}) appears due to the
electron tunneling from grain to grain, it was defined below
Eq.~(\ref{main}).

Deriving the analytical result for the diagrams in Figs.~1 and 2
it is important to take into account the fact that the single
electron propagator itself gets renormalized due to electron
hopping. Tunneling processes give rise an additional term to the
self-energy part of the single electron propagator, see Fig.~3
\begin{equation}
\label{self} \tau^{-1} = \tau_0^{-1} + 2  d g_T\delta ,
\end{equation}
where $\tau_0$ is the unrenormalized electron mean free time and
$g_T = 2\pi t^2\nu^2$ is the dimensionless tunneling conductance.
Although the second term in the right hand side of
Eq.~(\ref{self}) is much smaller than the first one it is
important to keep it because the leading order contribution in
$\tau_0^{-1}$ to the coupling constant $\lambda$ cancels ( see
Eqs.~(\ref{af}) and (\ref{ad}) below).

\begin{figure}[tbp]
\resizebox{.43\textwidth}{!}{\includegraphics{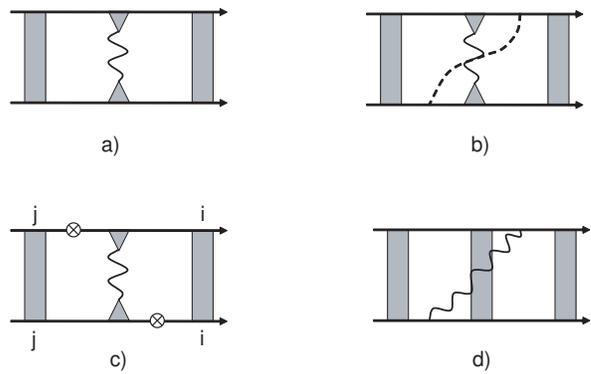}}
\vspace{0.6cm} \caption{Diagrams describing the first order vertex
corrections, $\delta\lambda_2$ to the  Cooperon channel coupling
constant $\lambda$ of granular metals due to the electron-electron
repulsion. All notations are the same as in Fig.~1. The diagrams
(b) - (d) should be taken with additional factor of $2$. After
that the final result for the sum of the diagrams (a) - (d) is
given by Eq.~(\ref{ad}).} \label{fig:3}
\end{figure}

We consider the contributions from diagrams in Fig.~1  and 2
separately. Using the result of Eq.~(\ref{self}) it is
straightforward to check that the sum of the diagrams (a)-(f) in
Fig.~1 results in the following correction, $\delta\lambda_1$ to
the Cooperon channel coupling constant
\begin{eqnarray}
\frac{\delta \lambda_1}{\lambda} = - \pi T^2 \sum\limits_{{\bf q}}
\left[ \sum\limits_{\varepsilon_n(\varepsilon_n -\Omega_n)<0}
\left( \frac{2V(\Omega_n, {\bf q})}{|\varepsilon_n|\,(|\Omega_n| +
\varepsilon_{\bf q}\delta)^2 }    \hspace{2cm}
\right. \right. \nonumber \\
\left. \left. + \frac{ V(\Omega, {\bf
q})}{\varepsilon_n^2(|\Omega_n| + \varepsilon_{\bf q}\delta
)}\right) - \sum\limits_{\varepsilon_n(\varepsilon_n -\Omega_n)>
0}\frac{V(\Omega_n, {\bf q})}{\varepsilon_n^2(|2 \varepsilon_n-
\Omega_n| + \varepsilon_{\bf q}\delta)} \right]. \hspace{1cm}
\label{af}
\end{eqnarray}
Here the summation is going over the quasi-momentum, ${\bf q}$ and
over the fermionic, $\varepsilon_n = \pi T(2n+1)$ and bosonic,
$\Omega_n = 2\pi T n$ Matsubara frequencies. The propagator of the
screened electron-electron interaction, $V(\Omega_n ,\mathbf{q})$
in Eq.~(\ref{af}) is given by the expression
\begin{equation}
V(\Omega_n ,\mathbf{q})={\frac{{2E_{C}(\mathbf{q})(|\Omega_n| +
\varepsilon_{\bf q}\delta )}}{{4\varepsilon
_{\mathbf{q}}E_{C}(\mathbf{q} )+ |\Omega_n| }}}. \label{V1}
\end{equation}
The parameter $\varepsilon _{\bf q}$ in Eqs.~(\ref{af}) and
(\ref{V1}) was defined below Eq.~(\ref{mainresult1}). The charging
energy $E_{C}(\mathbf{q})=e^{2}/2C(\mathbf{q})$ in Eq.~(\ref{V1})
is expressed in terms of the Fourier transform of the capacitance
matrix $C(\mathbf{q})$ which has the following asymptotic form
\begin{equation}
C^{-1}(\mathbf{q})=\frac{2}{a^D}\,\left\{
\begin{array}{rl}
& \pi /q\hspace{1.4cm} D=2, \\
& 2\pi /q^{2}\hspace{1.1cm} D=3.
\end{array}
\right.  \label{capacitance}
\end{equation}

\begin{figure}[htb]
\includegraphics[width=.75\linewidth]{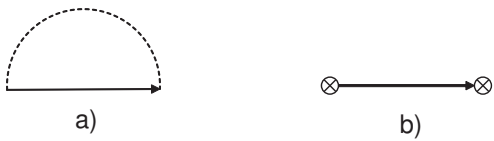}
\caption{Diagrams describing self-energy corrections to the single
electron propagator due to a) elastic interaction of electrons
with impurities and b) electron hopping. The solid lines denote
the bare propagator of electrons and the dashed line describes the
elastic interaction of electrons with impurities. The tunneling
vertices are described by the circles.} \label{fig:2}
\end{figure}

To obtain the total correction, $\delta\lambda$ to the Cooperon
channel coupling constant $\lambda$ due to electron-electron
interaction the diagrams in Fig.~2 that represent vertex
corrections should be taken into account. These diagrams results
in the following contribution
\begin{eqnarray}
\frac{\delta\lambda_2}{\lambda} = - \pi T^2 \sum\limits_{{\bf
q}}\left[ \sum\limits_{\varepsilon_n(\varepsilon_n -\Omega_n)<0}
\frac{V(\Omega_n, {\bf q})}{|\varepsilon_n||\varepsilon_n
-\Omega_n|(|\Omega_n| + \varepsilon_{\bf q}\delta)} \right. \nonumber \\
\left. + \sum\limits_{\varepsilon_n(\varepsilon_n -\Omega_n)>0}
\frac{V(\Omega_n, {\bf q})}{ |\varepsilon_n||\varepsilon_n
-\Omega_n|(|2\varepsilon_n - \Omega_n| + \varepsilon_{\bf q}\delta
)} \right] \label{ad}.
\end{eqnarray}
Using Eqs.~(\ref{V1}), (\ref{capacitance}) and summing over the
frequencies $\varepsilon_n$, $\Omega_n$ and quasi-momentum,
$\mathbf{q} $ in Eqs.~(\ref{af}) and (\ref{ad}) with the
logarithmic accuracy we obtain the final result (\ref{main}) for
the superconducting transition temperature in granular metals.

In conclusion, we considered the suppression of superconducting
transition temperature due to Coulomb repulsion in granular
metals. We found the correction to the transition temperature for
$3D$ granular samples and films at large tunneling conductance
between the grains. We demonstrated that the suppression of
superconductivity in $3D$ samples can be characterized by the
critical value of the conductance $g_T^*$ introduced in
Eq.~(\ref{g_star}) such that for samples with $g_T<g_T^*$ the
superconductivity is  suppressed. Taking into account effects of
Coulomb blockade of conductivity that become essential for samples
with $g_T<g_T^C,$ where the critical value $g_T^C$ is given by
Eq.~(\ref{gC}) we introduced the classification of $3D$ granular
samples composed from superconducting grains according to the
possibility of having insulating, metallic or superconducting
phases at zero temperature.

We would like to thank A.~I.~Larkin for useful discussions. This
work was supported by the U.S. Department of Energy, Office of
Science through contract No. W-31-109-ENG-38.

\end{document}